\documentclass[aps,twocolumn,superscriptaddress,prl,longbibliography]
{revtex4-1}

\usepackage{xcolor}
\usepackage{amsfonts}
\usepackage{amsmath}
\usepackage{dcolumn}
\usepackage{upgreek}
\usepackage{bm}
\usepackage{tikz}
\usepackage{ulem}
\usepackage[colorlinks=true,citecolor=blue,urlcolor=blue]{hyperref}
\usepackage{amsmath,amsfonts,amssymb,times}
\usepackage{natbib} 
\usepackage{jabbrv}
\usepackage[standard]{ntheorem}

\makeatletter
\newcommand*{\Rom}[1]{\expandafter\@slowromancap\romannumeral #1@}
\makeatother

\begin{document}

\title{Tunable Antichiral Hinge State in Photonic Synthetic Dimensions}

\author{Xian-Hao Wei}
\affiliation{Laboratory of Quantum Information, University of Science and Technology of China, Hefei 230026, China}
\affiliation{Anhui Province Key Laboratory of Quantum Network, University of Science and Technology of China, Hefei 230026, China}
\affiliation{CAS Center For Excellence in Quantum Information and Quantum Physics, University of Science and Technology of China, Hefei 230026, China}

\author{Xi-Wang Luo}\email{luoxw@ustc.edu.cn}
\affiliation{Laboratory of Quantum Information, University of Science and Technology of China, Hefei 230026, China}
\affiliation{Anhui Province Key Laboratory of Quantum Network, University of Science and Technology of China, Hefei 230026, China}
\affiliation{CAS Center For Excellence in Quantum Information and Quantum Physics, University of Science and Technology of China, Hefei 230026, China}
\affiliation{Hefei National Laboratory, University of Science and Technology of China, Hefei 230088, China}

\author{Mu Yang}
\affiliation{Laboratory of Quantum Information, University of Science and Technology of China, Hefei 230026, China}
\affiliation{Anhui Province Key Laboratory of Quantum Network, University of Science and Technology of China, Hefei 230026, China}
\affiliation{CAS Center For Excellence in Quantum Information and Quantum Physics, University of Science and Technology of China, Hefei 230026, China}

\author{Yu-Wei Liao}
\affiliation{Laboratory of Quantum Information, University of Science and Technology of China, Hefei 230026, China}
\affiliation{Anhui Province Key Laboratory of Quantum Network, University of Science and Technology of China, Hefei 230026, China}
\affiliation{CAS Center For Excellence in Quantum Information and Quantum Physics, University of Science and Technology of China, Hefei 230026, China}

\author{Jin-Shi Xu}
\affiliation{Laboratory of Quantum Information, University of Science and Technology of China, Hefei 230026, China}
\affiliation{Anhui Province Key Laboratory of Quantum Network, University of Science and Technology of China, Hefei 230026, China}
\affiliation{CAS Center For Excellence in Quantum Information and Quantum Physics, University of Science and Technology of China, Hefei 230026, China}
\affiliation{Hefei National Laboratory, University of Science and Technology of China, Hefei 230088, China}

\author{Guang-Can Guo}
\affiliation{Laboratory of Quantum Information, University of Science and Technology of China, Hefei 230026, China}
\affiliation{Anhui Province Key Laboratory of Quantum Network, University of Science and Technology of China, Hefei 230026, China}
\affiliation{CAS Center For Excellence in Quantum Information and Quantum Physics, University of Science and Technology of China, Hefei 230026, China}
\affiliation{Hefei National Laboratory, University of Science and Technology of China, Hefei 230088, China}

\author{Zheng-Wei Zhou}\email{zwzhou@ustc.edu.cn}
\affiliation{Laboratory of Quantum Information, University of Science and Technology of China, Hefei 230026, China}
\affiliation{Anhui Province Key Laboratory of Quantum Network, University of Science and Technology of China, Hefei 230026, China}
\affiliation{CAS Center For Excellence in Quantum Information and Quantum Physics, University of Science and Technology of China, Hefei 230026, China}
\affiliation{Hefei National Laboratory, University of Science and Technology of China, Hefei 230088, China}

\date{\today}

\begin{abstract}
Recent research in 2-dimensional (2D) topological matter has generalized the notion of edge states from chiral to antichiral configurations with the same propagating direction at parallel edges, revealing a rich variety of robust transport phenomena. Here, we propose that antichiral hinge states can emerge in a 3D higher-order topological insulator/semimetal, where two surface/bulk Dirac points are connected by the hinge states. The band dispersion can be controlled and tilted independently for each hinge using properly designed tunnelings, resulting in tunable antichiral hinge states with programmable propagation direction and velocity. Moreover, we propose experimental realization schemes based on a 1D coupled cavity array with additional synthetic dimensions represented by the photonic orbital angular momentum and frequency. We innovatively introduce both longitudinal and transversal electro-optic modulators 
to generate the desired tunable tunnelings along the synthetic dimensions, which significantly reduce the experimental complexity by eliminating the need for beam splittings and auxiliary cavities. The tunable antichiral hinge states are confirmed by the photonic transmission spectra. Our work presents the robust and tunable antichiral hinge-state transports which paves the way for exploring novel topological matter and  their device applications. 
\end{abstract}

\maketitle
\textit{\textcolor{blue}{Introduction}.---}The discovery and realization of novel topological phases of matter have emerged as one of the most vibrant research frontiers in contemporary physics~\cite{kanereview,shouchengreview}, attracting substantial interest from both fundamental science and technological applications. Initially identified in electronic systems, topological phases have since been successfully implemented in various synthetic quantum platforms including photonic~\cite{roadmapphotonics}, acoustic~\cite{secondwave}, and ultracold atomic systems~\cite{shiliangreview}.
A hallmark feature of topological matter is the emergence of robust one-way transport of edge modes, the paradigmatic example is quantum Hall systems, where chiral edge states propagate in opposite directions at parallel edges~\cite{haldanemodel,qwz}.
More remarkably, recent theoretical and experimental studies have generalized the concept of edge states from chiral to antichiral configurations based on a modified Haldane model, where edge states propagate in the same direction on the two opposite edges~\cite{antichiraloriginal} while maintaining backscattering immunity, this substantially increases the richness of robust unidirectional energy transport.
The concept of 1D chiral edge state is further extended to 2D antichiral surface states~\cite{antichiral3dweylprx,antichiral3dweylnjp,antichiral3d1,antichiral3d2}. Since their initial theoretical proposal, antichiral boundary modes have been experimentally realized in multiple physical platforms~\cite{antichiraltwist,antichiralferro,antichiralpolariton1,antichiralpolariton2,antichiralcircuit,antichiralopticaltheo2,antichiralopticaltheo3,antichiralopticalexp1,antichiralopticalexp3,antichiralopticalexp4,antichiralopticalexp6,antichiralopticalexp7,antichiralopticalexp8}.

\begin{figure}[t]
    \includegraphics[width=0.6\linewidth]{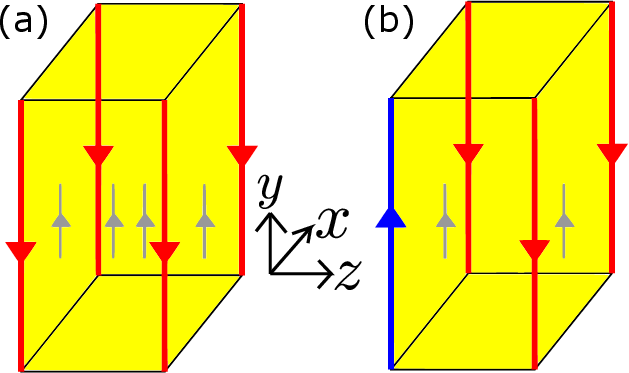}
    \caption{Schematic of tunable antichiral hinge state. (a) The four hinge modes propagate along the same direction (red arrows). (b) One of the hinge modes is tuned to reverse its propagation direction (blue arrow). The thin gray arrows represent the currents of surface or bulk modes compensating the hinge currents.}
    \label{newanti1}
\end{figure}

Meanwhile, recent years have witnessed the discovery of higher-order topological insulators in which a $d$-dimensional $n$-th order system hosts robust $(d-n)$-dimensional boundary modes protected by bulk topology~\cite{higherorderscience,higherorderprb,higherordernrp,higherorderxureview,higherorderscienceadvances,higherorderdirac,higherorderzhuyanqing,higherorderxiwang,higherorderfangchen,higherorderweyljiang,higherorderhybrid,higherorderwallpaper,higherordernonsymm,higherorder170page,higherorderexp20241,higherorderezawa1,higherorderliuzheng,hotianti1,hotianti2,higherorderparityanomaly,higherorderexpextra1,higherorderexpextra2,higherorderexpextra3,higherorderldos}. For example, chiral hinge modes can emerge in a 3D second-order topological insulator which may facilitate wider transport manipulation capabilities than 2D topological architectures.
Therefore, integrating higher-order topological hinge states with antichiral configurations would enable more versatile energy transport devices with unprecedented functionalities, which remains unexplored though. 
A recent theoretical attempt of Ref.~\cite{antichiralstack2d} relies on stacking two high-order topological insulators with opposite chiral hinge states, which can form antichiral boundary states at two hinges. However, it is still unclear how to obtain antichiral hinge states within a single higher-order topological bulk, as shown in Fig.~\ref{newanti1}(a), where complex tunnelings are expected for its realization. On the application side, especially in the realm of photonics, tunability and reconfigurability of the topological boundary modes are usually preferred [see Fig.~\ref{newanti1}(b)], the photonic crystal usually has limited tunability after fabrication. 
In this context, recent advances in engineering photonic synthetic dimensions~\cite{photosynreview1,photosynreview2,photosynreview3,luo2015,luo2017,yuantwosynthetic,xiangfaopen,photoexp1,photoexp5,photoexp6,photoexp7,photoexp8,photoexp9,photoexp11,photoexp12,photoexp13,photoexp14,photoexp15,photoexp16,photoexp19,photoexp22,4dfloquet,higherordersyntheticdimension1,higherordersyntheticdimension2,nonabelianelectric,tensor2,syntheticdong}, where unprecedented parameter tunability (e.g., lattice geometries, tunnelings) and dynamic reconfiguration are possible, offer new opportunities in realizing the aforementioned topological phases and transports, which demand the study.

In this work, we propose a 3D four-band topological lattice model supporting four tunable antichiral hinge state and provide its experimental realization scheme based on synthetic dimensions. 
Depending on the presence or absence of the $C_4$ symmetry in the $x$-$y$ plane, the system can be either a Dirac semimetal or insulator (with gapless or gapped bulk in the Brillouin zone), with tilted hinge (along $z$ direction) Fermi arcs connecting two Dirac points in the bulk or surface, respectively. 
The Fermi arcs can be controlled and tilted independently for each hinge using properly designed tunnelings, resulting in tunable antichiral hinge states with programmable propagation direction and velocity.
The hinge currents are compensated by counter-propagating bulk or surface modes to ensure zero total current, as shown in Figs.~\ref{newanti1}(a) (b).
Experimentally, we consider the realization of our model in photonic systems based on two synthetic dimensions represented by angular momentum (both orbital and polarization) and frequency~\cite{yuantwosynthetic,higherordersyntheticdimension2}, in addition to one real-space dimension with coupled cavity arrays. 
Along the synthetic dimensions,
different angular momentum states (representing $y$-direction lattice sites) can be coupled by q-plate~\cite{qplate} and wave-plate~\cite{photoexp14,photoexp15,photoexp16}, while different frequency modes (representing $z$-direction lattice sites) can be coupled by electro-optics modulator (EOM) which can be dynamically controlled by the applied electric fields.
Both longitudinal and transversal EOMs are employed to couple different frequency and polarization modes. These couplings along synthetic dimensions need neither auxiliary cavities nor beam splittings, thereby simplify experimental apparatus significantly. 
The couplings along real-space dimension ($x$ direction) can realized using auxiliary cavities. It is worthy to note that EOMs with sawtooth-modulated electric field or acousto-optic modulators (AOMs) are used in the auxiliary cavities to generate diagonal tunnelings along the real and frequency dimensions (i.e., additionally 
desired spin-orbit couplings).
To probe the antichiral hinge modes, we consider a finite real-space lattice length, and create a sharp boundary in the angular-momentum dimension by inserting a phase plate (PP) with a pinhole into the cavity,
then the hinge state propagation can be observed based on the transmission spectra.

\begin{figure*}[t]
    \includegraphics[width=1.0\linewidth]{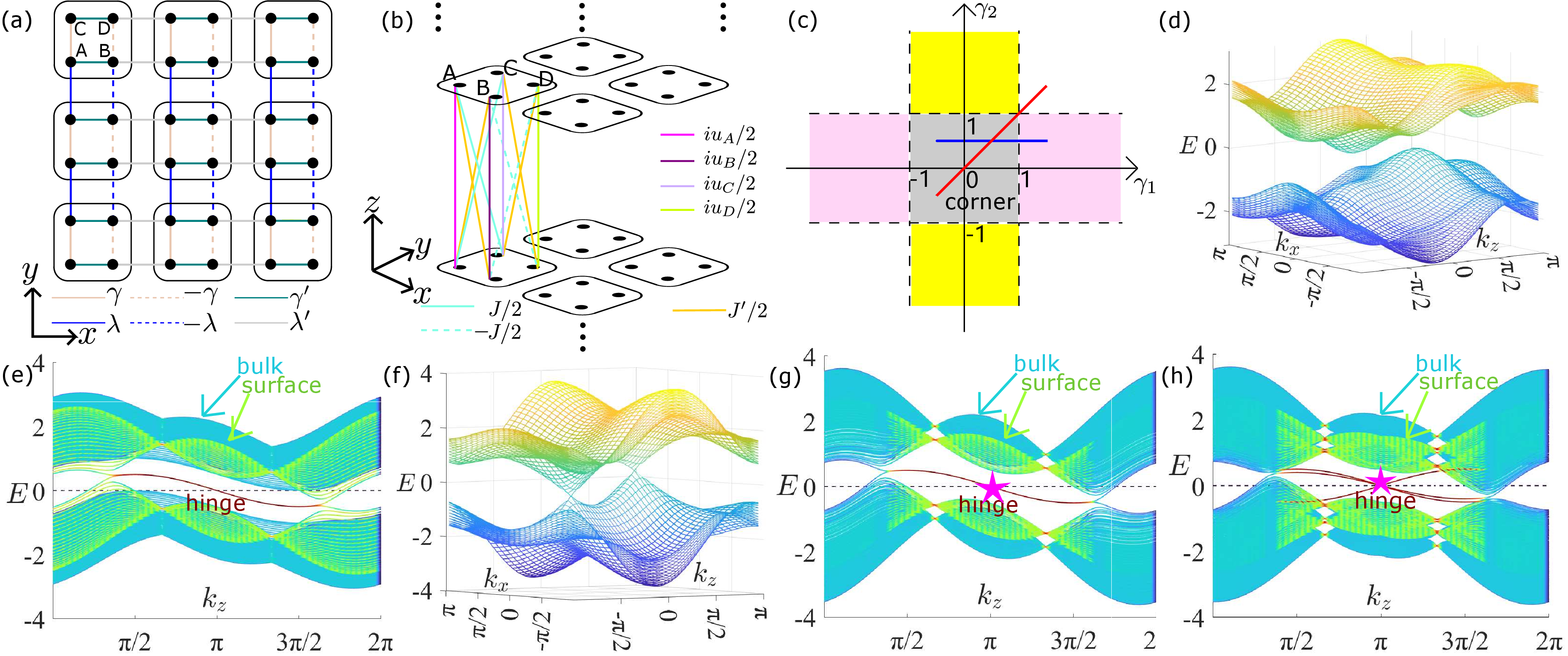}
    \caption{(a) and (b) Schematic of the lattice model Eq.~(\ref{2dhoti}) and Eq.~(\ref{3dtsm}), respectively. 
    In (b), only the tunnelings along $z$ are shown for one unit cell. The corresponding tunneling amplitudes are also denoted in the plot. (c) Phase diagram of the reduced 2D model by fixing $k_z$ in Eq.~\ref{3dtsm}. The blue (red) line correspond to the path of $\gamma_1(k_z)$-$\gamma_2(k_z)$ when $k_z$ varies from 0 to $2\pi$ for $J'=0$ ($J'=J$), where $\gamma_1(k_z)=\gamma+J\cos k_z,\gamma_2(k_z)=\gamma'+J'\cos k_z$.
    (d) Energy spectrum of Eq.~(\ref{3dtsm}) with periodic boundary conditions along all directions for $J'=0$ and $u_s=0.5$ $\forall s$. (e) Same as that in (d) but with open boundary conditions along $x$ and $y$ directions. Besides the hinge modes, there are also surface modes, only the surface modes at $x=0,N_x$ are shown, see Appendix B for the surface modes at $y=0,N_y$. (f) and (g) Same as that in (d) and (e) but with $J'=J$.
    (h) Same as that in (g) but with reconfigured tiling $u_A=-0.5$, $u_B=u_D=0.5$ and $u_C=0.4$.
    We see the direction and group velocity of the antichiral hinge modes are tuned accordingly. Common parameters are $\gamma=\gamma'=0.5$, with system size $N_x=N_y=24,N_z=60$ and energy unit $-\lambda=-\lambda'=J=1$.
    }
    \label{newanti2}
\end{figure*}

\begin{figure*}[t]
    \centering
    \includegraphics[width=0.75\linewidth]{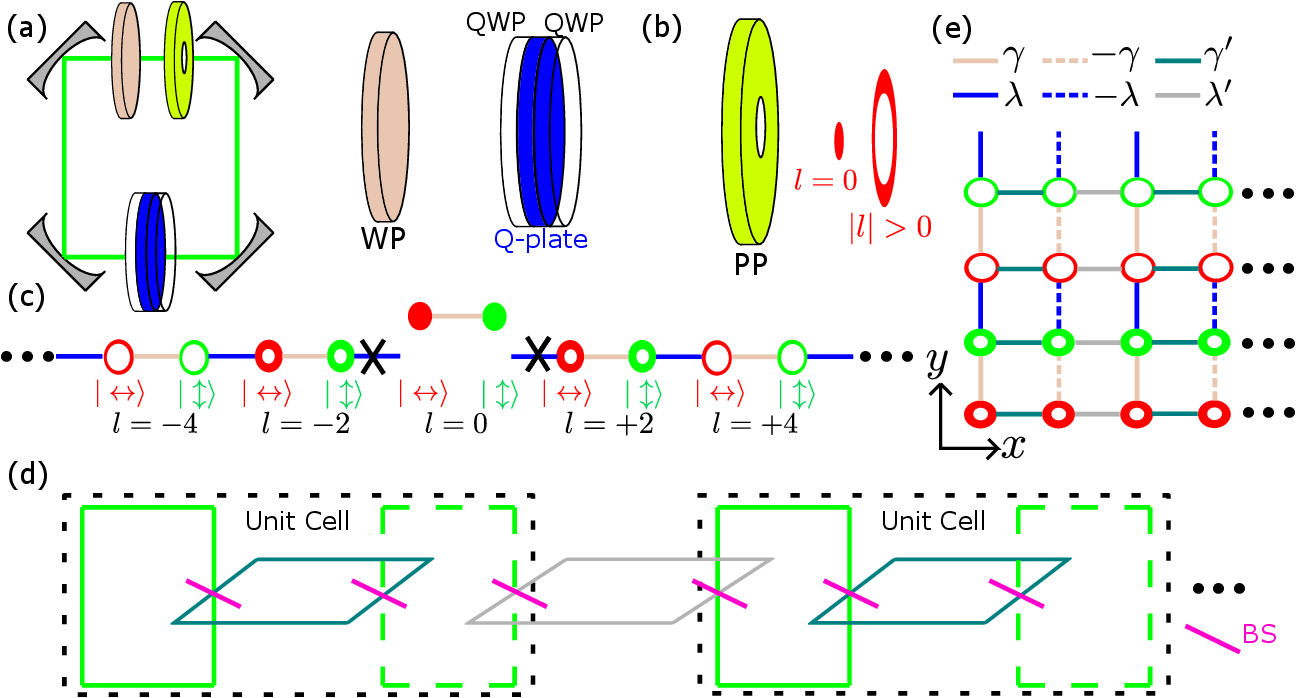}
    \caption{Schematic of experimental setup to realize the 2D lattice model Eq.~(\ref{2dhoti}). (a) A single degenerate cavity with optical elements coupling the synthetic angular momentum dimension. (b) Phase plate with a pinhole used to create the sharp boundary along the synthetic dimension. The schematic of mode density distributions for $l=0$ and $|l|>0$ are also shown. (c) The lattice structure corresponding to the setup in (a), a sharp boundary is formed at $l=0$. (d) 1D coupled cavity arrays by coupling the degenerate cavity shown in (a). There are two subcavities in each unit cell, they have opposite tunneling sign along the synthetic angular momentum dimension. BS represents the beam splitter. (e) The lattice structure of 2D lattice model Eq.~(\ref{2dhoti}) corresponding to the setup in (d).    
   }
    \label{newanti3}
\end{figure*}

\textit{\textcolor{blue}{Model}.---}To start with, we first introduce the 2-dimensional second-order topological insulator supporting topological-protected corner states. Similar as the well-known BBH model~\cite{higherorderscience,higherorderprb}, the lattice configuration is shown in Fig.~\ref{newanti2}(a), with four sublattice sites ($A,B,C,D$) in each unit cell, 
the Bloch Hamiltonian reads
\begin{equation}
    \begin{split}
        H_\text{2D}(k_x,k_y)=&(\gamma+\lambda\cos k_y)\Gamma_4+\lambda\sin k_y\Gamma_3\\
        &+(\gamma'+\lambda'\cos k_x)\Gamma_2+\lambda'\sin k_x\Gamma_1
    \end{split}
    \label{2dhoti}
\end{equation}
Where $\Gamma_1=\sigma_0\tau_y,\Gamma_2=\sigma_0\tau_x,\Gamma_3=\sigma_y\tau_z,\Gamma_4=\sigma_x\tau_z$ obey the Clifford algebra $\{\Gamma_1,\Gamma_2\}=2\delta_{ij}$, $\sigma_i$ and $\tau_i$ are the Pauli matrix representing the sublattice sites in $y$ and $x$ directions, respectively, $\sigma_0$ is the identity matrix. The system enters the topological phase with four corner states in the parameter region $|\gamma|<|\lambda|$ and $|\gamma'|<|\lambda'|$. 

Next, we introduce the third direction by stacking the above 2D system along $z$ and including inter-layer couplings.
The 3D lattice tunneling configuration is shown in Fig.~\ref{newanti2}(b) with Bloch Hamiltonian given by
\begin{align}
    H_\text{3D}(\textbf{k})=&(\gamma_1(k_z)+\lambda\cos k_y)\Gamma_4+\lambda\sin k_y\Gamma_3\nonumber\\
    +&(\gamma_2(k_z)+\lambda'\cos k_x)\Gamma_2+\lambda'\sin k_x\Gamma_1+u\sin k_z\nonumber\\
    =&\sum_jh_j(\textbf{k})\Gamma_j+H_u\label{3dtsm}
\end{align}
Where $\gamma_1(k_z)=\gamma+J\cos k_z$ and $\gamma_2(k_z)=\gamma'+J'\cos k_z$. Here $J$ is the tunneling strength along $z$ with sublattice site flipping $A\leftrightarrow C$ and $B\leftrightarrow D$, while
$J'$ is the tunneling strength along $z$ with sublattice site flipping $A\leftrightarrow B$ and $C\leftrightarrow D$, as shown in Fig.~\ref{newanti2}(b).
$H_u=u\sin k_z\sigma_0\tau_0$ is the pseudo-potential corresponding to tunnelings along $z$ with a $\pi/2$ phase but without sublattice site flipping,
we first consider the tunnelings that are independent of sublattice sites. 
With these couplings, the corner states evolve into hinge states along $z$ in the region  $|\gamma_1(k_z)|<|\lambda|$ and $|\gamma_2(k_z)|<|\lambda'|$, while the pseudo-potential term $H_u$ tilts the hinge-mode dispersion
and leads to the so-called antichiral hinge modes with one-way propagation.
The eigenenergies of the bulk modes are $E_{\pm}=\pm\sqrt{\sum_jh_j(\textbf{k})^2}+u\sin k_z$, each with twofold degeneracy. The bulk gap (at each Bloch momentum in the Brillouin zone) closes only if $h_j(\textbf{k})=0$ $\forall j$, otherwise the bulk state is fully gapped. 
When $u=0$, the Hamiltonian Eq.~(\ref{3dtsm}) has reflection symmetry: $M_iH_\text{3D}(\textbf{k})M_i=H_\text{3D}(-k_i,k_{j\neq i})$, where $M_x=\sigma_z\tau_x,M_y=\sigma_x,M_z=I$, as well as time-reversal and chiral symmetry: $\mathcal{K}H_\text{3D}(\textbf{k})\mathcal{K}=H_\text{3D}(-\textbf{k}),\;\Xi H_\text{3D}(\textbf{k})\Xi^{-1}=-H_\text{3D}(\textbf{k})$, where $\Xi=\sigma_z\tau_z$ and $\mathcal{K}$ corresponds to the complex conjugation operator. When $u\neq0$, the pseudo-potential term breaks time-reversal symmetry and reflection symmetry along $z$, but preserves their combination \textcolor{red}{$\Xi\mathcal{K}$}. 
For simplicity, we will set $-1<\gamma'=\gamma<1,\lambda'=\lambda=-J=-1$ across this paper unless otherwise stated. With these choices of parameters,
the Hamiltonian also has rotational symmetry in the $x$-$y$ plane: $C_4^{xy}H_\text{3D}(k_x,k_y,k_z)(C_4^{xy})^{-1}=H_\text{3D}(-k_y,k_x,k_z)$ when $J'=J$, where $C_4^{xy}=\exp(-i\frac{\pi}{4}\sigma_y\tau_x)$. 

For a given $k_z$, the Hamiltonian $H_\text{3D}$ is reduced to an effective 2D system, whose phase diagram in the $\gamma_1(k_z)$-$\gamma_2(k_z)$ plane are shown in Fig.~\ref{newanti2}(c). Surface modes at the boundary of $x$ (or $y$) exist in the region $|\gamma_2|<|\lambda'|$ (or $|\gamma_1|<|\lambda|$), while hinge modes (i.e., corner modes in 2D) exist in the region
$|\gamma_2|<|\lambda'|$, $|\gamma_1|<|\lambda|$.
As $k_z$ changes from $0$ to $2\pi$,
the trajectory of $(\gamma_1,\gamma_2)$ forms a closed path, as shown by the solid  blue ($J'=0$) and red  $(J'=J)$ lines in the phase diagram Fig.~\ref{newanti2}(c).
Notice that, the bulk gap closes only at the point $\gamma_1(k_z)=-\lambda$, $\gamma_2(k_z)=-\lambda'$. Therefore, for $J'=0$, the path of $(\gamma_1,\gamma_2)$ does not cross the bulk-gap closing point, the Hamiltonian $H_\text{3D}$ is a higher-order topological insulator with gapped bulk, as shown in Fig.~\ref{newanti2}(d). While for $J'=J$, the path of $(\gamma_1,\gamma_2)$ crosses the bulk-gap closing point, the system becomes a higher-order topological semimetal with two bulk Dirac points, as shown in Fig.~\ref{newanti2}(f).
We also notice that, the path of $(\gamma_1,\gamma_2)$ partially lies in the region with corner modes, indicating the existence of hinge states for certain $k_z$ intervals.
In the following, we will solve for these hinge states and their band dispersions.

\textit{\textcolor{blue}{Antichiral hinge states}.---}We can solve for the effective hinge states using a similar method as in Ref.~\cite{higherorderexact2}. 
To do so, we consider semi-infinite boundary conditions along  $x,y$ directions and periodic boundary condition along $z$ direction. Due to the reflection symmetry in $x$ and $y$ directions,
we will consider the hinge at $x=0$, $y=0$ only and solve for the hinge modes by assuming the ansatz:
\begin{equation}
|\psi_\text{hinge}\rangle=\sum_{x,y\geq0}\mathcal{N}b(k_z)^yc(k_z)^x|x,y\rangle\otimes|\xi\rangle.
\end{equation}
Where $\mathcal{N}$ is normalization factor, and $|\xi\rangle$ is sublattice wave function. By solving the equation $H|\psi_\text{hinge}\rangle=E_\text{h}|\psi_\text{hinge}\rangle$, we can obtain the solution $|\xi\rangle=\left|A\right\rangle$ with $b(k_z)=-\gamma_1(k_z)/\lambda,\;c(k_z)=-\gamma_2(k_z)/\lambda'$, with eigenenergy of hinge modes  $E_\text{h}=u\sin k_z$ (see Appendix A for more details).
The existence of hinge mode solution requires that $|b(k_z)|,|c(k_z)|<1$. 
Here, we focus on $J'=0$ and $J'=J$, as depicted by the paths of $[\gamma_1(k_z),\gamma_2(k_z)]$ when $k_z$ varies from 0 to $2\pi$ in Fig.~\ref{newanti2}(c), we can see that the hinge modes lie in the interval $\cos^{-1}(1-\gamma)<k_z<2\pi-\cos^{-1}(1-\gamma)$. 
By considering periodic boundary conditions along both $z$ and $y$ ($x$), we can solve for the surface states at $x=0$ ($y=0$) in a similar way (see Appendix A).

Numerically, we can solve the system by considering open boundary conditions along $x,y$ directions and periodic boundary condition along $z$ direction.
When $J'=0$, the bulk is fully gapped and the hinge Fermi arcs connect the Dirac point of the surface modes ($x=0$ and $x=N_x$ with $N_x$ the length along $x$), as shown in Fig.~\ref{newanti2}(e). The pseudo-potential $H_u=u\sin k_z$ tilts the band dispersion of the hinge states, 
for the Fermi energy $E_F=0$ as dashed line in Fig.~\ref{newanti2}(e), the four hinge modes acquire the same group velocity and propagation direction, exhibiting  antichiral feature. The upward and downward currents in any finite lattice systems must be the same, where bulk or surface modes will compensate the hinge current, as shown in Fig.~\ref{newanti1}. From Fig.~\ref{newanti2}(e), we can see that the compensation current is mainly carried by the surface states. 
For $J'=J$, the system obeys $C_4^{xy}$ symmetry, and the bulk states close the gap at certain $k_z$, as shown in 
Fig.~\ref{newanti2}(g), the hinge Fermi arcs connect Dirac point of the bulk states, and the compensation current is mainly carried by the bulk states. We want to point out that, as we increase the tilting (i.e., group velocity of antichiral hinge modes) with a larger $u$, the Fermi energy $E_F=0$ will cut both the surface and bulk modes, the compensation current will be carried by both the bulk and surface states. 
Moreover, when $J'>J=-\lambda$, we find the tilted hinge Fermi arcs split into two intervals in $k_z$ (see Appendix B for more details).

As we discussed above, for a fixed $k_z$, the system reduced to a 2D model similar to the BBH model, the hinge state is reduced to the corner state, its topological origin are given by the quadrupole $Q_{xy}(k_z)$ in the $x$-$y$ space, which
can be obtained by the Wannier polarization or edge polarization through the nested Wilson loop theory~\cite{higherorderscience,higherorderprb} (see Appendix B for more details).
Notice that the pseudo-potential $H_u$ corresponds to a $k_z$ dependent energy offset which doesn't change the topological number.

In order to tune the propagation direction and velocity of the hinge modes, we first note that each hinge mode is mainly distributed on some particular sublattice sites. Therefore, we can introduce sublattice-site dependent pseudo-potential to tilt the band dispersion of each hinge mode independently. That is, we can replace $H_u$ by the following $H'_u$
\begin{eqnarray}
    H'_u=\sum_{s={A,B,C,D}}u_s \sin(k_z) |s\rangle\langle s|
\end{eqnarray}
The sign and amplitude of the tunneling parameter $u_s$ control the direction and velocity of the hinge mode on sublattice site $|s\rangle$. This can be verified by solving the hinge modes based on the method discussed above, which would lead to the hinge dispersion $E_s=u_s \sin(k_z)$. As an example, we flip the propagation direction of hinge mode on sublattice $A$ by setting $u_B=u_D=-u_A=0.5$ and $u_C=0.4$, the corresponding band structures are shown in Fig.~\ref{newanti2}(h).

\begin{figure*}[t]
    \centering
    \includegraphics[width=0.8\linewidth]{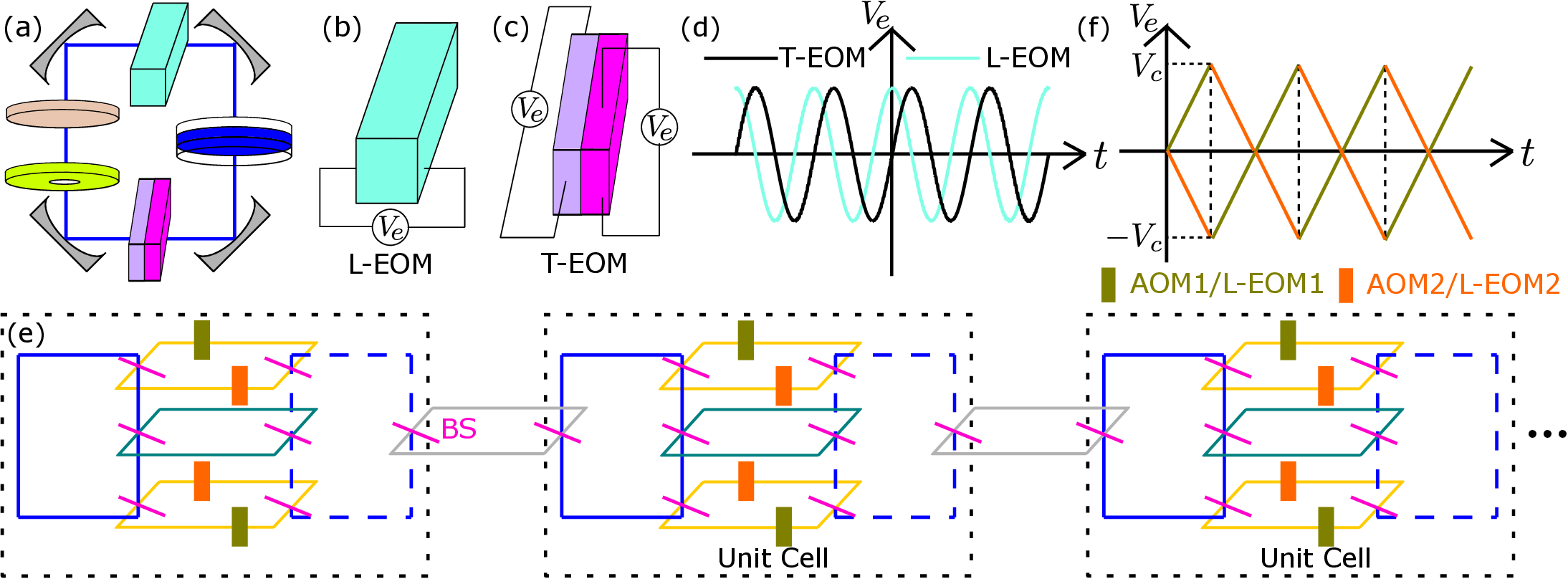}
    \caption{Schematic of experimental setup to realize the 3D lattice model Eq.~(\ref{3dtsm}). (a) A single degenerate cavity with two additional EOMs (L-EOM and T-EOM) compared with Fig.~\ref{newanti3}(a). By applying suitable electric field to the two EOMs as shown in (b) and (c), desired couplings along the synthetic frequency dimension can be realized. (d) The waveform of the applied electric fields to the EOMs in (b) and (c). (e) 1D coupled cavity arrays by coupling the degenerate cavity shown in (a). Two additional auxiliary cavities are introduced within each unit cell compared with Fig.~\ref{newanti2}(d)
    which are used to generate the diagonal tunnelings in the $x$-$z$ plane. Each arm of the two additional auxiliary cavities contains an AOM or EOM which shifts the frequency of photons passing through by $\pm\Omega_\text{FSR}$. (f) The waveform of the applied electric fields for the EOMs in the auxiliary cavities in (e).  
    }
    \label{newanti4}
\end{figure*}

\textit{\textcolor{blue}{Experimental realization scheme}.---}We propose an experimental implementation scheme based on a 
1D coupled cavity array for $x$ direction with two additional sythetic dimensions represented by orbital angular momentum (OAM) and frequency for $y$ and $z$ directions.
First, we show how to realize the 2D Hamiltonian $H_\text{2D}$ in the $x$-$y$ space.
We consider the degenerate optical cavities supporting different OAM modes~\cite{degeneratecavity1,degeneratecavity2,chengexp,chengexp2,chengexp3}. The cavities also support different frequency modes with splitting of free spectral range (FSR) $\Omega_\text{FSR}=2\pi\frac{L}{c}$, with $L$ the cavity path length and $c$ the speed of light. The four sublattice sites in each unit cell are represented by two polarization states and two sub-cavities.
Along the synthetic $y$ direction, the unit cell index is represented by OAM number, while the sublattice index is represented by the polarization (horizontal $\left|\leftrightarrow\right\rangle$ and vertical $\left|\updownarrow\right\rangle$). The intra-cell coupling (i.e., $\gamma$) can be realized by simply adding a wave-plate (WP) inside the cavity [as shown in Fig.~\ref{newanti3}(a)], which couple the two polarization states without affecting the OAM states~\cite{photoexp14}.
While the inter-cell coupling (i.e., $\lambda$) is implemented by a q-plate~\cite{qplate,photoexp14} inside the cavity [see Fig.~\ref{newanti3}(a)]. When the light passes through the q-plate, a small portion of left-handed (right-handed) circularly polarized light experiences $+2$ ($-2$) OAM shift accompanied by changing to right-handed (right-handed), see Appendix C for more details. Since the sublattice sites are represented by the two linear polarization states, we sandwich the q-plate between two quarter wave plates (QWPs) with optical axis at $45^\circ$ with respect to the horizontal axis, so the QWPs will transform the basis from circular polarization to linear polarization. Sharp boundaries in this synthetic dimension can be constructed using a phase plate (PP) with a pinhole at the center, as shown in Figs.~\ref{newanti3}(a) (b),
where $l=0$ mode is the Gaussian mode with most intensity distributed within the center pinhole while $l\neq0$ modes all have a torus-like shape with negligible distributions within the center pinhole~\cite{xiangfaopen,photoexp16}.
Such a PP would generate a different phase delay for $l=0$ mode compared with all $l\neq0$ modes, leading to a large energy offset which decouples the $l=0$ mode from the chain. These optical elements generate a 1D synthetic lattices with open boundaries shown in Fig.~\ref{newanti3}(c).

Along the real-space dimension $x$, the sublattice sites are represented by two subcavities, the intra-cell (i.e., $\gamma'$) and inter-cell (i.e., $\lambda'$) couplings can both be realized via auxiliary cavities~\cite{luo2015}. {The auxiliary cavities are intentionally operated in an off-resonant regime, so that most photons stays inside in subcavities}. The tunnelling phase and amplitude can be controlled independently for each site, as shown in Fig.~\ref{newanti3}(d). Notice that, to generate the $\pi$ flux at each plaquette, the tunnellings along the $y$ synthetic dimension for the right subcavity have opposite sign compared to the left subcavity, this can be realized by rotating the WP by $90^\circ$ and changing the thickness of the 
q-plate for the right subcavity. With all the coupling along $x$ and $y$ directions, we obtain the 2D Hamiltonian with lattice configuration shown in Fig.~\ref{newanti3}(e) [same as that in Fig.~\ref{newanti2}(a)].

\begin{figure*}[t]
    \centering
    \includegraphics[width=1.0\linewidth]{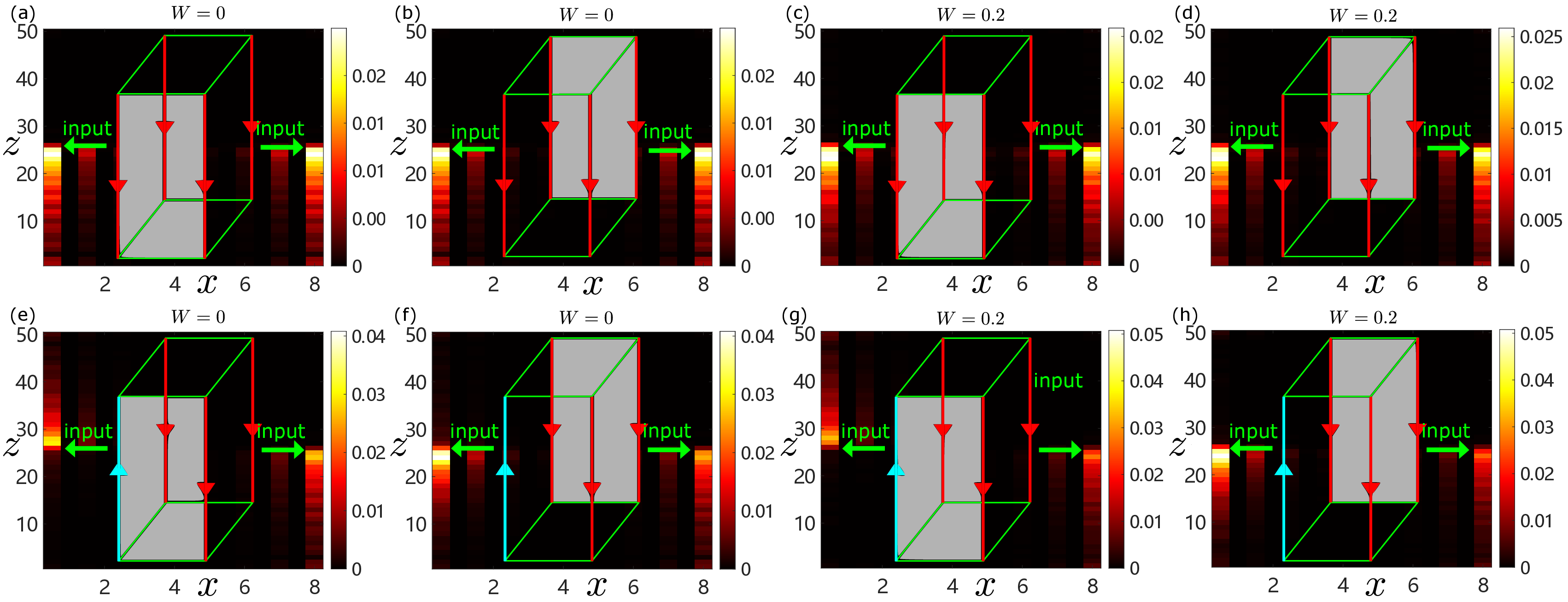}
    \caption{Detecting the transports of tunable antichiral hinge states. We excite three sites in a row at each hinge with delayed phase $k_z^*$ (Eq.~\ref{psiin}) and quasi-energy $E=0$.
    (a) and (b) The transmissions at the surface $y=0$ and $y=N_y$, respectively, with other parameters the same as that in Fig.~\ref{newanti2}(g).
    (c) and (d) Same as that in (a) and (b) but with {on-site} disorders, the disorder strength is $W=0.2$, the plots are averaged over 50 realizations. (e) and (f) The transmissions at the surface $y=0$ and $y=N_y$, respectively, with other parameters the same as that in Fig.~\ref{newanti2}(h). We can see that the transport direction for the hinge on sublattice site $A$ is reversed and the transport distance for the hinge on sublattice site $C$ is reduced.
    (g) and (h) Same as that in (e) and (f) but with disorders, the disorder strength is $W=0.2$, the plots are averaged over 50 realizations. Common parameters: $E_\text{in}=1$ for all plots.    
    }
    \label{newanti5}
\end{figure*}

Next, we show how the couplings along $z$ direction (represented by different frequencies separated by $\Omega_\text{FSR}$) can be realized experimentally. Different from 
previous spin-orbit coupling schemes which utilize polarization beam splitter to separate the polarization states and modulate them independently~\cite{nonabelianelectric}, here we use polarization-dependent EOMs with longitudinal and transversal electro-optics modulation to realize the desired spin-orbit coupling, without the need of beam splittings. 
We notice that there are three types of couplings: $J\cos (k_z)\Gamma_4$, $J'\cos (k_z) \Gamma_2$ and $u_s \sin (k_z) |s\rangle\langle s|$.

(i) The coupling $J\cos (k_z)\Gamma_4$ corresponds to tunnelings along $z$ direction with polarization flip
$A\leftrightarrow C$ and $B\leftrightarrow D$ while the subcavity index remain unchanged, as shown in  Fig.~\ref{newanti2}(b). To realize this coupling, we insert a crystal like KDP with $\bar{4}2$ symmetry inside the cavity and apply an electric field $E_{\bar{z}}$ along the $\bar{z}$-axis, as shown in Fig.~\ref{newanti4}(a) (b). Such a longitudinal electro-optics modulator (denoted as L-EOM) has the following polarization-dependent reflection index~\cite{eom1,eom2}
\begin{equation}
    n_{\bar{x}}\approx n_0+n_0^3\frac{\gamma_{63}}{2}E_{\bar{z}},n_{\bar{y}}\approx n_0-n_0^3\frac{\gamma_{63}}{2}E_{\bar{z}},n_{\bar{z}}=n_e
\label{nxxyyz}
\end{equation}
Where $\bar{x}$ and $\bar{y}$ span the polarization plane while $\bar{z}$ represents the propagation direction of light. We apply electric field by the voltage $V_{e}(t)={V_m}\cos(\Omega_{\text{FSR}}t+\varphi)$ [see Fig.~\ref{newanti4}(d)], and the two polarization states feel opposite phase modulations. We set $\bar{x}$, $\bar{y}$ direction being rotated by $45^\circ$ with respect to the horizontal and vertical polarization directions, the polarization basis is rotated and
the Jone matrix of the L-EOM is
\begin{equation}
U_\text{L-EOM}=Se^{-iGV_c\cos(\Omega_\text{FSR}t+\varphi)\sigma_z}S^{\dagger}
\label{szs}
\end{equation}
with basis rotation operator
\begin{equation}
S=\frac{1}{\sqrt{2}}\left[
\begin{array}{cccc}
1&-1\\
1&1\\
\end{array}
\right]
\end{equation}
Where $G=\frac{1}{2}n_0^3\gamma_{63}\frac{\omega_0}{c}$, where $\omega_0$ is the central frequency of the light.
For $GV_m\ll 1$, Eq.~(\ref{szs}) can be approximated as $I-\frac{iGV_m}{2}e^{i\varphi}e^{i\Omega_\text{FSR}t}\sigma_x-\frac{iGV_m}{2}e^{-i\varphi}e^{-i\Omega_\text{FSR}t}\sigma_x$, leading to the coupling between different frequencies $\omega\rightarrow\omega\mp\Omega_\text{FSR}$ accompanied by polarization flipping. 
It related with the Hamiltonian as $U_\text{L-EOM}=e^{-iH_\text{L-EOM}T_R}$ (where $T_R=L/c$ denotes the round-trip time of photon inside the cavity), resulting in $H_{\text{L-EOM}}=J\cos (k_z+\varphi)\sigma_x$ with $J=\frac{GV_m\Omega_\text{FSR}}{2\pi}$ and $k_z=\Omega_\text{FSR}t$.
Notice that for the synthetic frequency lattice, $t$ is the reciprocal quantity of frequency and can be interpreted as the wavevector $k_z$, the eigenenergy $E$ is interpreted as the Floquet quasi-energy of the system.
For the left subcavity, we apply the electric field with $\varphi=0$ to realized the coupling $J\cos (k_z)\sigma_x$, for the right subcavity, we apply the electric field with $\varphi=\pi$ to realized the coupling $-J\cos (k_z)\sigma_x$, together they give rise to the desired coupling $J\cos (k_z)\Gamma_4$.

(ii) The coupling $J'\cos (k_z)\Gamma_2$ term
corresponds to tunnelings along $z$ direction with subcavity flip
$A\leftrightarrow B$ and $C\leftrightarrow D$ while the polarization remains unchanged, as shown in  Fig.~\ref{newanti2}(b). This can be realized by  diagonal (next-nearest-neighbor) tunnelings in the $x$-$z$ space (real and frequency dimensions).
We can introduce two additional auxiliary cavities to couple the two subcavities with each unit cell,
and insert an optical element in each arm of the auxiliary cavity that shifts photons' frequency by $\Omega_\text{FSR}$, the modulation in the two arms of each auxiliary cavity compensate with each other.
The frequency shifting element can be a high-efficiency AOM~\cite{PhysRevApplied.23.014031} or a
longitudinal modulated EOM, as shown in Fig.~\ref{newanti4}(e). For the EOM, the two axis $\bar{x},\bar{y}$ should be parallel with the two polarization basis, and sawtooth voltage modulation with slope $\zeta$ and maximum $V_c$ should be used, as shown in Fig.~\ref{newanti4}(f). The EOM in one arm induces the phase modulation as
$e^{\pm i \int G\zeta(t) dt}$ for the two polarization states. We set the slope such that $G|\zeta|=\Omega_\text{FSR}$ and $GV_c=n\pi$ so that the phase modulation can be replaced by
$e^{\pm i \Omega_\text{FSR} t}$ (up to a $2n\pi$ phase). The auxiliary cavity
leads to the coupling $(\omega,A)\leftrightarrow(\omega+\Omega_\text{FSR},B)$ and
$(\omega,C)\leftrightarrow(\omega-\Omega_\text{FSR},D)$. Similarly, a second auxiliary cavity can be designed to generate the coupling
$(\omega,A)\leftrightarrow(\omega-\Omega_\text{FSR},B)$ and
$(\omega,C)\leftrightarrow(\omega+\Omega_\text{FSR},D)$, together they realize the desired term $J'\cos k_z\Gamma_2$, with $J'$ determined by the beam splitters connecting the auxiliary cavity and subcavity. 


(iii) To realize the term $u_s \sin(k_z)|s\rangle\langle s|$, we place two EOMs inside the cavity and apply transversal electric field, as shown in Figs.~\ref{newanti4}(a) (c). The two transverse EOMs (denoted as T-EOMs) modulate the phase delay of horizontal and vertical polarization, respectively.
For the left subcavity, the Jones matrix of the two T-EOMs is $e^{-iGV_{A}\cos(\Omega_\text{FSR}t+\varphi_A)}|A\rangle\langle A| +e^{-iGV_{C}\cos(\Omega_\text{FSR}t+\varphi_C)}|C\rangle\langle C|$, with
$V_A$, $V_C$ ($\varphi_A$, $\varphi_C$) the amplitudes (phases) of the applied electric fields. 
Similar results can be obtained for the right subcavity. The corresponding coupling can be written as $u_s \sin(k_z+\varphi_s+\pi/2)|s\rangle\langle s|$ with
$u_s=GV_{s}/T_R$, see Appendix C for more details. It is clear that the tilting potential $H'_u$ can be realized and tuned dynamically by the modulation amplitudes and phases of the applied electric fields [see Fig.~\ref{newanti4}(d)].

\textit{\textcolor{blue}{Detecting Method}.---}We propose a detecting method based on the transimision spectra to observe tunable antichiral hinge states. The input-output relation of the system is~\cite{luo2015,photoexp1}
\begin{equation}
    \Psi_\text{out}(E)=(I-\frac{i\kappa}{E-H+i\kappa/2})\Psi_\text{in}(E)=(I+\mathbb{T})\Psi_\text{in}
\end{equation}
Where $\Psi_\text{in}$ and $\Psi_\text{out}$ are the states of input and output optical fields.
We focus on the transmission spectrum $\Psi_\text{T,out}=\mathbb{T}\Psi_\text{in}$ term. We want to excite the hinge mode with quasi-energy and momentum $(E,k_z^*)$ around the pink star point in Fig.~\ref{newanti2}, this
can be done by pumping the three sites $(z_c-1,z_c,z_c+1)$ in a row at the hinge, with delayed phase $k_z^*$ between two sites. The input field can be set as
\begin{equation}
    \begin{split}
        \Psi_\text{in}=&\frac{E_\text{in}}{\sqrt{3}}|x_\text{in},y_\text{in}\rangle\otimes(|z_c+1\rangle\\
        &+e^{ik_z^*}|z_c\rangle+e^{2ik_z^*}|z_c-1\rangle)
    \end{split}
    \label{psiin}
\end{equation}
Where $x_\text{in}=0,N_x$ and $y_\text{in}=0,N_y$ denote corner sites in $xy$-plane.
$E_\text{in}$ denotes the input optical field. 
The transmission field can be written as 
\begin{equation}
    \Psi_\text{T,out}=\sum_{k_z}\frac{-i\kappa E_\text{in}\alpha(k_z,k_z^{*})}{E-H(k_z)+i\kappa/2}|x_\text{in},y_\text{in}\rangle\otimes|k_z\rangle
\end{equation}
Where $\alpha(k_z,k_z^*)=(1+e^{i(k_z^*-k_z)}+e^{2i(k_z^*-k_z)})$, and $H(k_z)$ is Hamiltonian with open (periodic) boundary conditions along $x,y$ ($z$). We note that the output intensity is proportional to $\alpha$ which has a sharp peak around $k_z=k_z^*$. We also set the input-output coupling rate $\kappa$ much smaller than the energy gap near $k_z^*$, such that
the transmission is mostly contributed by the hinge modes.

In Fig.~\ref{newanti5}, we numerically calculated the output light intensity $|\mathbb{T}\Psi_\text{in}|^2$ for different hinges with different tilting term $H'_u$, the effects of disorders are also investigated. In Fig.~\ref{newanti5}(a)-(d) with $u_s=u$ $\forall s$, we find the antichiral light transports with the four hinges propagating along the same direction. 
Such transports are robust against the on-site disorders $\sum_iW_ic_i^{\dagger}c_i$ with $W_i\in[-W,W]$. 
Similarly, in Fig.~\ref{newanti5}(e)-(h) we flip the sign of $u_A$ and find the reversed hinge propagation along the hinge of $A$ sublattice site which is robust against disorders, indicating the tunability of the  propagation direction and velocity of the anti-chiral hinge states. 

\textit{\textcolor{blue}{Conclusion}.---}In summary, 
we have proposed a 3D lattice model supporting four independently tunable antichiral hinge states and provided the experimental realization schemes utilizing photonic synthetic dimensions.
The topological bulk (insulator or semimetal) is controlled by $C_4$ symmetry, manifesting as hinge Fermi arcs bridging surface or bulk Dirac points, respectively. 
Crucially, the independently tilted dispersion of each hinge mode enables on-demand antichiral transport, which is achieved via our innovative electro-optic control scheme combining longitudinal/transverse modulators and sawtooth-modulated electric field---eliminating the need for beam splitting and reducing the number of auxiliary cavities and thereby enhancing experimental feasibility. The antichiral dynamics are directly observable through the transmission spectra, establishing this platform as a versatile tool for designing robust photonic devices. Our work paves the way for exploring higher-order topology and engineering robust photonic transport in synthetic-lattice systems.

\begin{widetext}
\setcounter{figure}{0} \renewcommand{\thefigure}{A\arabic{figure}} %
\setcounter{equation}{0} \renewcommand{\theequation}{A\arabic{equation}}

\appendix

\section{Appendix A: Hinge and surface states}
As we discussed in the main text, the hinge states at
$x=0,y=0$ can be solved for according to the 
Schrodinger equation $H|\psi_\text{hinge}\rangle=E|\psi_\text{hinge}\rangle$, with
the ansatz
\begin{equation}    |\psi_\text{hinge}\rangle=\sum_{x,y\geq0}\mathcal{N}b(k_z)^yc(k_z)^x|x,y\rangle\otimes|\xi\rangle
\end{equation}
Where $H$ is the Hamiltonian with open boundaries along $x,y$ directions and periodic boundary along $z$, which reads
\begin{equation}
    \begin{split}
        H=&(\gamma_1(k_z)+\lambda\frac{|x,y+1\rangle\langle x,y|+|x,y\rangle\langle x,y+1|}{2})\Gamma_4+\lambda\frac{|x,y+1\rangle\langle x,y|-|x,y\rangle\langle x,y+1|}{2i}\Gamma_3\\
        &+(\gamma_2(k_z)+\lambda'\frac{|x+1,y\rangle\langle x,y|+|x,y\rangle\langle x+1,y|}{2})\Gamma_2+\lambda'\frac{|x+1,y\rangle\langle x,y|-|x,y\rangle\langle x+1,y|}{2i}\Gamma_1+u\sin k_z
    \end{split}
\end{equation}
With in the bulk $x>0,y>0$, we have
\begin{equation}
    \begin{split}
        [&(\gamma_1(k_z)+\lambda\frac{b(k_z)+b(k_z)^{-1}}{2})\Gamma_4+\lambda\frac{b(k_z)-b(k_z)^{-1}}{2i}\Gamma_3+(\gamma_2(k_z)+\lambda'\frac{c(k_z)+c(k_z)^{-1}}{2})\Gamma_2\\
        &+\lambda'\frac{c(k_z)-c(k_z)^{-1}}{2i}\Gamma_1]|\xi\rangle\otimes|x,y\rangle=(E-u\sin k_z)|\xi\rangle\otimes|x,y\rangle
        \label{bulkeff}
    \end{split}
\end{equation}
The boundary condition at the $x=0,y=0$ becomes
\begin{equation}
    \begin{split}
        [&(\gamma_1(k_z)+\lambda\frac{b(k_z)}{2})\Gamma_4+\lambda\frac{b(k_z)}{2i}\Gamma_3+(\gamma_2(k_z)+\lambda'\frac{c(k_z)}{2})\Gamma_2\\
        &+\lambda'\frac{c(k_z)}{2i}\Gamma_1]|\xi\rangle\otimes|0,0\rangle=(E-u\sin k_z)|\xi\rangle\otimes|0,0\rangle
        \label{cornereff}
    \end{split}
\end{equation}
The boundary condition at the $x>0,y=0$ becomes
\begin{equation}
    \begin{split}
        [&(\gamma(k_z)+\lambda\frac{b(k_z)+b(k_z)^{-1}}{2})\Gamma_4+\lambda\frac{b(k_z)-b(k_z)^{-1}}{2i}\Gamma_3+(\gamma_2(k_z)+\lambda'\frac{c(k_z)}{2})\Gamma_2\\
        &+\lambda'\frac{c(k_z)}{2i}\Gamma_1]|\xi\rangle\otimes|x,0\rangle=(E-u\sin k_z)|\xi\rangle\otimes|x,0\rangle
        \label{edgeeff}
    \end{split}
\end{equation}
The boundary condition at the $x=0,y>0$ becomes
\begin{equation}
    \begin{split}
        [&(\gamma_1(k_z)+\lambda\frac{b(k_z)}{2})\Gamma_4+\lambda\frac{b(k_z)}{2i}\Gamma_3+(\gamma_2(k_z)+\lambda'\frac{c(k_z)+c(k_z)^{-1}}{2})\Gamma_2\\
        &+\lambda'\frac{c(k_z)-c(k_z)^{-1}}{2i}\Gamma_1]|\xi\rangle\otimes|0,y\rangle=(E-u\sin k_z)|\xi\rangle\otimes|0,y\rangle
        \label{edgeeff2}
    \end{split}
\end{equation}
From these equations, we find $|\xi\rangle$ satisfies 
$P_{x;0}|\xi\rangle=|\xi\rangle$ and $P_{y;0}|\xi\rangle=|\xi\rangle$, with the projection operators
\begin{equation}
    P_{x;0}=\frac{I+i\Gamma_1\Gamma_2}{2},P_{y;0}=\frac{I+i\Gamma_3\Gamma_4}{2}
\end{equation}
which lead to $|\xi\rangle=|A\rangle$.
Also, we can deduce that $\gamma_1(k_z)+\lambda b(k_z)=\gamma_2(k_z)+\lambda'c(k_z)=0,\;E=u\sin k_z$. The existence of hinge state requires that $|b(k_z)|,|c(k_z)|<1$, that is $|\gamma_1(k_z)|<|\lambda|,|\gamma_2(k_z)|<|\lambda'|$.
Notice that the hinge states at $x=0,y=0$ only occupy the sublattice sites $A$, the solution will not be affected by repacing $H_u$ by $H'_u$. In this case, the hinge state energy becomes $E=u_A\sin(k_z)$.
Similarly, we can solve for the hinge states on other hinges, the total hinge state Hamiltonian reads 
\begin{equation}
    H_\text{hinge}=\sum_s u_s\sin(k_z) |\psi_{\text{hinge},s}\rangle\langle \psi_{\text{hinge},s}|
\end{equation}
As a result, the band dispersion for each hinge can be tuned independently.

We can also solve for the surface states in a similar way. For example, we consider the surface at $x=0$ and set periodic boundaries along both $y$ and $z$. The ansatz of surface state at $x=0$ is:
\begin{equation}    |\psi_{x,\text{surface}}\rangle=\sum_{x\geq0}\mathcal{N}c(k_y,k_z)^x|x\rangle\otimes|\xi'\rangle
\end{equation}
We find that $|\xi'\rangle$ satifies
$P_{x;0}|\xi'\rangle=|\xi'\rangle$ {with subspace} $\{|\xi'\rangle\}=\{|A\rangle,|B\rangle\}$. {The existence of surface states requires $|c(k_y,k_z)|<1$, that is $|\gamma_2(k_z)|<|\lambda'|$.}
With in the interval of $k_z$ satisfying $|\gamma_2(k_z)|<|\lambda'|$, we can project the Hamiltonian in the surface state subspace to obtain the effective surface Hamiltonian at $x=0$ as
\begin{equation}
    H_{x,\text{surface}}(k_y,k_z)=P_{x;0}HP_{x;0}=[\gamma_1(k_z)+\lambda\cos (k_y)]\sigma_x+\lambda\sin k_y\sigma_y+u\sin k_z
\end{equation}
Similar results can be found for other surfaces.

\begin{figure}
    \centering
    \includegraphics[width=0.9\textwidth]{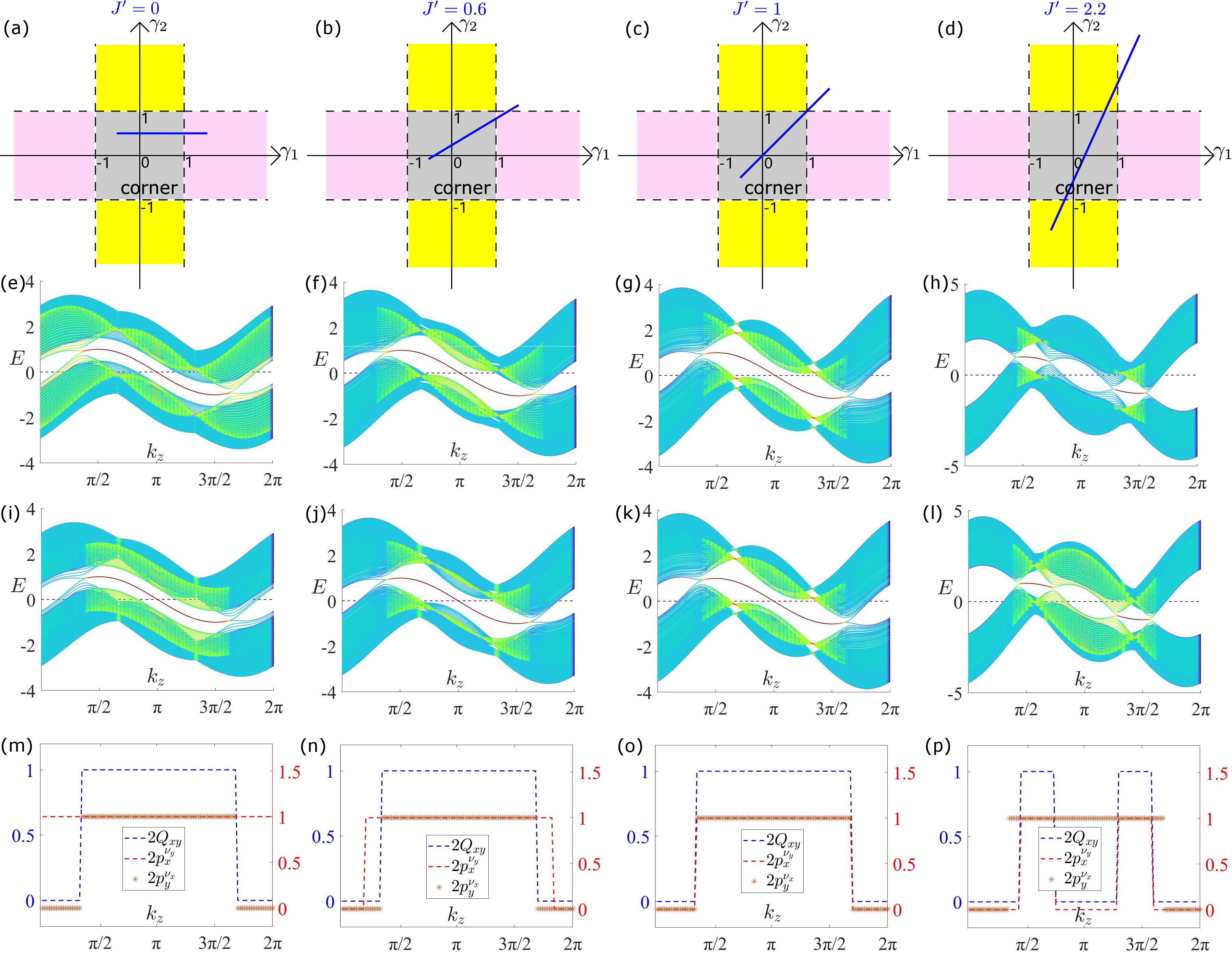}
    \caption{(a)-(d) Phase diagram of the reduced 2D system and the paths (blue lines) of $[\gamma_1(k_z),\gamma_2(k_z)]$ as $k_z$ varying from $0$ to $2\pi$ for the 3D Hamiltonian Eq.~(2) in the main text, with
$J'=0$, $J'=0.6$, $J'=1$ and $J'=2.2$, respectively. 
(e)-(h) Band structures corresponding to the paths in (a)-(d), with surface states on $x=0,N_x$ displayed. (i)-(l) Same as that in (e)-(h) but with surface states on $y=0,N_y$ displayed. In (e)-(l), we adopt open boundary conditions along $x,y$ and periodic boundary conditions along $z$. (m)-(p) Topological invariants $Q_{xy}(k_z),\;p_x^{\nu_y}(k_z),\;p_y^{\nu_x}(k_z)$ corresponding to the paths (a)-(d). Common parameters are: $\lambda=\lambda'=-J=-1,\gamma=\gamma'=0.5,u_A=u_B=u_C=u_D=1,N_x=N_y=24,N_z=60$.}
    \label{suppdifferentband}
\end{figure}

\section{Appendix B: Band structures and topological invariant}
\label{energybandtotal}
Here we show the band structures of our 3D lattice model in Eq.~(2) in the main text with $J'$ taking different values.
In Figs.~\ref{suppdifferentband} (a)-(d), the paths of $[\gamma_1(k_z),\gamma_2(k_z)]$ as $k_z$ varying from $0$ to $2\pi$ are shown by the blue lines with
$J'=0$, $J'=0.6$, $J'=1$ and $J'=2.2$, respectively.
The corresponding band structures are shown in Figs.~\ref{suppdifferentband} (e)-(h) with surface states on $x=0,N_x$ displayed and Figs.~\ref{suppdifferentband} (i)-(l)
with surface states on $y=0,N_y$ displayed, we have adopted open boundary conditions along $x,y$ and periodic boundary conditions along $z$.
We see that the hinge modes exist in the gap of bulk and surface modes.
When $0\leq J'\leq 1$, the hinge mode lies in the range $\cos^{-1}(1-\gamma)<k_z<2\pi-\cos^{-1}(1-\gamma)$. When $J'>1$, the hinge modes split into two intervals: $\cos^{-1}(\frac{1-\gamma}{J'})<k_z<\cos^{-1}(\frac{-1-\gamma}{J'})$ and $2\pi-\cos^{-1}(\frac{-1-\gamma}{J'})<k_z<2\pi-\cos^{-1}\frac{1-\gamma}{J'}$.

In Fig.~\ref{suppdifferentband}(e,f,i,j), we can see that the hinge Fermi arcs connect Dirac point in the surface ($x=0,N_x$). In Fig.~\ref{suppdifferentband}(h,l), we can see that the hinge Fermi arcs connect Dirac point in the surface ($y=0,N_y$). In Fig.~\ref{suppdifferentband}(g,k), we can see that the hinge Fermi arcs connect Dirac point in the bulk. More general paths can be considered in a similar way by choosing different forms of $\gamma_1(k_z)$ and 
$\gamma_2(k_z)$.

The topology of our system is characterized by the quadruple $Q_{xy}(k_z)$ of the reduced 2D model with fixed $k_z$, which determines the existence of hinge modes. Consider the Hamiltonian with uniform tilting $H_u$ so the system possesses reflection symmetry along $x$ and $y$, the topological invariant reduced 2D model can be obtained by nested Wilson loop method discussed in Ref.~\cite{higherorderscience,higherorderprb}. By calculating the Wilson loop of the occupied bands along $k_y$, one can obtain the Wannier bands given by $\pm\nu_{y}(k_z)$; then one can construct the Wannier basis according to one Wannier sector $\nu_y(k_z)$, and calculate the Wannier polarization $p^{\nu_y}_x(k_z)$ along $x$. Similarly, one can obtain the Wannier polarization along $y$ as $p^{\nu_x}_y(k_z)$. The topological invariant $Q_{xy}(k_z)=2p^{\nu_x}_y(k_z)p^{\nu_y}_x(k_z)$. The procedure is standard and we give the results directly here,
which are shown in Fig.~\ref{suppdifferentband}(m)-(p). We can see that at certain $k_z$, $Q_{xy}(k_z)=\frac{1}{2}$ indicating the existence of hinge states inside the gap of surface or bulk modes.
while for $Q_{xy}(k_z)=0$, no hinge states exist.
The existence of surface states can be determined by the Wannier polarization $p_x^{\nu_y}(k_z),p_y^{\nu_x}(k_z)$. 
We can see that $p_x^{\nu_y}=\frac{1}{2}$ ($p_y^{\nu_x}=\frac{1}{2}$) indicating the presence of surface states at $x=0,N_x$ ($y=0,N_y$), while $p_x^{\nu_y}=0$ ($p_y^{\nu_x}=0$) indicating the absence of surface states.

Finally, we note that when we consider the tunable antichiral hinge state with $H'_u$, the reflection symmetries are broken, the above topological characterization breaks down. However, for a large system such that the coupling between the four hinges can be omitted, the existence of hinge modes will not be affected but with modified band dispersion, as discussed in Appendix A.

\section{Appendix C: Experimental realization schemes}
Here we provide more details about the experimental realization schemes. First we show how the q-plate induces the couplings along angular momentum dimension. A q-plate is a set of liquid crystal molecules with different optical axes in a quasi-2D plane, each of which can be viewed as a wave plate~\cite{qplate}, it couples the left-handed (right-handed) circularly polarized photon with
right-handed (left-handed) while changing the OAM by
$-2q$ ($+2q$). We can set $q=1$, as a result, within the subspace
$\{|l+2,\circlearrowleft\rangle,|l,\circlearrowright\rangle\}$,
the q-plate leads to the transformation (up to a global phase delay)
\begin{eqnarray}
U_\text{q}=\cos(\frac{\chi}{2}) I -i \sin(\frac{\chi}{2})\sigma_x=e^{-i\chi \sigma_x}
\end{eqnarray}
with $\chi$ controlled by the thickness of the q-plate. When $\chi$ is small, the corresponding effective Hamiltonian 
can be obtained through $U_q=e^{-iH_q T_R}$, leading to $H_q=\frac{\chi}{2T_R} \sigma_x$. We can sandwich the q-plate between two QWPs to change the polarization basis from $\{|\circlearrowleft\rangle,|\circlearrowright\rangle\}$ to $\{|\leftrightarrow\rangle,|\updownarrow\rangle\}$. Considering all OAM states, the Hamiltonian of the composite q-plate can be written as
\begin{eqnarray}
H'_q=\frac{\chi}{2T_R}\sum_l |l,\updownarrow\rangle\left\langle l+2,\leftrightarrow\right| +h.c.   
\end{eqnarray}
achieving the targeted tunnelings. If we change $\chi\rightarrow \chi+\pi$, the sign of the tunneling is reversed.

Next, we provide more details about the method to realize tunable tilting term
$H'_u$.
Consider the left subcavity, we insert two EOMs with transversal electro-optics modulation, such EOMs are denoted as T-EOMs. We set the light propagation direction along $\bar{z}$, and horizontal $|\leftrightarrow\rangle$ (vertical $|\updownarrow\rangle$) polarization along the $\bar{x}$ ($\bar{y}$) direction.
The first T-EOM is used to modulate the phase delay of
the horizontal polarization $\leftrightarrow$ (i.e., A sublattice site) while leaving 
the vertical polarization $\updownarrow$ (i.e., C sublattice site) unchanged,
with electric field $E^A_{\bar{y}}$ applied along $\bar{y}$ direction, its  reflection
index is
\begin{equation}
    n^A_{\bar{x}}=n_0+n_0^3\frac{\gamma_{63}}{2}E^A_{\bar{y}},\;n^A_{\bar{y}}=n_e,\;n^A_{\bar{z}}=n_0-n_0^3\frac{\gamma_{63}}{2}E^A_{\bar{y}}
\end{equation}
We set $\frac{\omega_0}{c}L_{\bar{z}}(n_0-n_e)=2n\pi$, and the applied electric field  $E^A_{\bar{y}}(t)=\frac{V_A}{L_{\bar{y}}}\cos(\Omega_\text{FSR}t+\varphi_A)$, so that 
the Jones matrix reads
\begin{equation}
   U^A_T= |A\rangle\langle A|e^{-iGV_A\cos(\Omega_\text{FSR}t+\varphi_A)}+|C\rangle\langle C|
\end{equation}
$G=\frac{1}{2}n_0^3\gamma_{63}\frac{\omega_0}{c}\frac{L_{\bar{z}}}{L_{\bar{y}}}$, with $L_{\bar{y}}$ and $L_{\bar{z}}$ the thickness of the EOM and we can set $L_{\bar{y}}=L_{\bar{z}}$ for simplicity. Similarly, we can obtain the 
Jones matrix $U^C_T$ for the second T-EOM, together they give rise to 
\begin{equation}
   U^C_TU^A_T= |A\rangle\langle A|e^{-iGV_A\cos(\Omega_\text{FSR}t+\varphi_A)}+|C\rangle\langle C|e^{-iGV_C\cos(\Omega_\text{FSR}t+\varphi_C)}
\end{equation}
When $GV_A$ and $GV_C$ are small, we can
define the effecitve Hamiltonian through
$U^C_TU^A_T=e^{-iH_T^{AC}T_R}$, 
and $H_T^{AC}=u_A \sin(k_z+\varphi_A+\pi/2)|A\rangle\langle A|+u_C \sin(k_z+\varphi_C+\pi/2)|C\rangle\langle C|$
with
$u_s=GV_{s}/T_R$. Notice that time can be interpreted
as the wavevector $k_z$ for the synthetic frequency lattice.
Similar results can be obtained for the right subcavity {corresponding to $B$, $D$ sublattice sites}, and the independent tunable tilting for each hinge is achieved.

\end{widetext}

\textit{Acknowledgments.---}We thank Jin-Ming Cui and Ze-Di Cheng for useful discussions. This work is supported by the Natural Science Foundation of China (Grants No. 12474366, and No. 11974334) and Innovation Program for Quantum Science and Technology (Grants No. 2021ZD0301200). X.-W. Luo also acknowledges support from the USTC start-up funding.

%

\end{document}